\documentclass[a4paper,11pt]{article}
\usepackage{jheppub} % for details on the use of the package, please see the JINST-author-manual
\usepackage{lineno} %\linenumbers
\usepackage{appendix}
\usepackage{orcidlink}
\usepackage{hyperref}
\usepackage{mathtools}
\usepackage{bm}
\usepackage{soul}

\usepackage{comment}

\usepackage{mdfra med}
\newcommand{\be}{\begin{equation}}
\newcommand{\ee}{\end{equation}}
\definecolor{purpleheart}{rgb}{0.41, 0.21, 0.61}

\newcommand{\uD}{\text{D}}
\newcommand{\ud}{\,\mathrm{d}} %for infinitesimals i.e. \ud {t}

\newcommand{\appropto}{\mathrel{
    \vcenter{
        \offinterlineskip
        \halign{
            \hfil$##$\cr\propto\cr\noalign{\kern1pt}\sim\cr\noalign{\kern-1pt}
        }
    }
}}

\makeatletter
\gdef\@fpheader{} %uncomment to remove "prepared for submission to JCAP"
\makeatother

\preprint{APCTP Pre2026-005,  NORDITA-2026-029}

\title{Ultra slow-turn inflation}

\author[a]{Ana Ach\'ucarro,}
\affiliation[a]{Instituut-Lorentz of Theoretical Physics, Universiteit Leiden, 2333 CA Leiden, The Netherlands}
\emailAdd{achucar@lorentz.leidenuniv.nl}

\author[b]{Perseas Christodoulidis,}
\affiliation[b]{Department of Science Education, Ewha Womans University, Seoul 03760, Korea}
\emailAdd{perseas@ewha.ac.kr}

\author[b,c]{Jinn-Ouk Gong,}
\affiliation[c]{Asia Pacific Center for Theoretical Physics, Pohang 37673, Korea}
\emailAdd{jgong@ewha.ac.kr}

\author[d,e,a]{and Oksana Iarygina}
\affiliation[d]{Nordita, KTH Royal Institute of Technology and Stockholm University, %Hannes Alfv\'ens v\"ag 12, 
10691 Stockholm, Sweden}
\affiliation[e]{The Oskar Klein Centre, Stockholm University, 10691 Stockholm, Sweden}
\emailAdd{oksana.iarygina@su.se}

\abstract{ 
In standard multi-field models, tachyonic isocurvature perturbations generally indicate the presence of an instability. We revisit the stability of some known counterexamples and show that, in a certain class of models that we call \textit{ultra slow-turn}, an exponentially decreasing turn rate can shut off this potential instability. 
We argue that the stability of a given model can be correctly inferred by the total entropy perturbation, even if the effective mass squared of the isocurvature perturbation is negative.
Several recent supergravity- or string-inspired models such as fibre inflation, SL(2,$\mathbb{Z}$) attractors and modular inflation fall into the ultra slow-turn class.}

\begin{document}
\maketitle
\flushbottom

\section{Introduction}

%Inflation, why attractor behaviour is necessary, etc..
The leading candidate to explain the observed homogeneity and isotropy of the cosmic microwave background (CMB) is an early phase of accelerated expansion of the universe -- cosmic inflation (see \cite{Achucarro:2022qrl, Ellis:2023wic,Kallosh:2025ijd} for recent reviews). Another important role of inflation is that the tiny initial density perturbations responsible for the temperature anisotropies of the CMB and the large-scale distribution of galaxies are seeded during inflation via amplifying vacuum quantum fluctuations. The most recent observations of the CMB are, with high accuracy, consistent with the anticipated properties of the primordial perturbation \cite{Planck:2018jri,AtacamaCosmologyTelescope:2025blo}. One crucial property is that the amplitude of the primordial perturbation is nearly scale-invariant. This means that, at least for the scales relevant for the CMB observations, the dynamics of the primordial perturbation has become nearly independent from the initial conditions of various $k$-modes so that the amplitudes of these modes converge to a stable constant value (see e.g.~\cite{Mukhanov:2005sc,Baumann:2022mni}). This ``attractor'' behaviour is an important feature of cosmological perturbations produced during inflation.

The attractor solution can be straightforwardly obtained if inflation is driven by a single inflaton field. To implement inflation we require the potential to be dominant over the kinetic energy of the inflaton. This ``slow-roll'' condition in turn allows us to neglect the second time derivative of the inflaton, leading to a first-order differential equation of motion. This means the phase space for the dynamics of the inflaton is collapsed to one-dimension, with momentum being irrelevant, so the attractor behaviour is exhibited stably during the slow-roll phase. Meanwhile, as the inflaton candidate is absent in the Standard Model of particle physics (SM), the realization of inflation usually requires theories beyond the SM or effective field theory. In these theories, there are multiple scalar degrees of freedom that can be dynamically relevant at the inflationary energy, as high as $10^{15}$ GeV. Thus, it is natural to presume that during inflation multiple degrees of freedom actively participate in dynamic -- multi-field inflation. For reviews see e.g.~\cite{Malik:2008im,Langlois:2010xc,Gong:2016qmq}. The analysis of the stable attractor, however, becomes subtle and demands more care.

After the seminal work of Gordon et al.~\cite{Gordon:2000hv} the orthonormal frame became the main tool to study multi-field models \cite{GrootNibbelink:2000vx,GrootNibbelink:2001qt}. In this basis, perturbations acquire a physical meaning: The projection tangential to the trajectory is related to the adiabatic perturbation, whereas those along the orthogonal directions are related to entropy or non-adiabatic perturbations. The limit $k\rightarrow 0$ of the perturbation equations has become the consensus to investigate the stability of a background trajectory. In this limit, the orthogonal perturbation decouples from the adiabatic one and satisfies a single-field equation with mass $\mu_{\rm eff}^2$, therefore requiring $\mu_{\rm eff}^2>0$ for stability. This matches the behaviour of most models; in the slow-turn case, the background becomes destabilized whenever $\mu_{\rm eff}^2$ becomes negative, for instance
if the potential becomes tachyonic or whenever the field space curvature becomes large and negative (with representative examples, the hybrid inflation model \cite{Linde:1993cn} and the geometrical destabilization scenario \cite{Gong:2011uw,Renaux-Petel:2015mga}, respectively).

However, a class of models was presented in~\cite{Cicoli:2018ccr}  with shift symmetry in one of the two fields, in both the potential and the field metric, which seem to defy the requirement $\mu_{\rm eff}^2>0$. More specifically, it was shown  that although the effective mass is negative, the background solution is perfectly stable \cite{Cicoli:2019ulk}. Meanwhile, Christodoulidis et al.~\cite{Christodoulidis:2019mkj,Christodoulidis:2022vww} directly linearized the background equations in a fairly general class of models and demonstrated that the stability criteria do not necessarily involve the effective mass, giving a partial answer to this apparent paradox. Later, Cicoli et al.~\cite{Cicoli:2021yhb,Cicoli:2021itv} returned to the issue of tachyonic isocurvature perturbations and argued that these are irrelevant because they are not directly observable and advocated for the use of the relative entropy perturbation, which was shown to remain finite. Recently, there has been renewed interest in models with $\mu_{\rm eff}^2<0$ motivated by supergravity \cite{Kallosh:2024ymt,Kallosh:2024pat,Kallosh:2024kgt,Carrasco:2025rud,GonzalezQuaglia:2025qem}. In these models, stability is inferred in the field basis, where both fields perturbations remain finite. Moreover, these models typically predict single-field evolution, and, therefore, the effect of isocurvature perturbations is typically ignored, as a vanishing turn rate does not feed the curvature perturbation on super-horizon scales.

In this work, we suggest to focus on the behaviour of the curvature and \textit{total entropy} perturbations as a diagnostic tool for stability and argue that these two variables are the relevant ones from both a \textit{physical and a mathematical} point of view.\footnote{%This was missed in the analysis of \cite{Cicoli:2021yhb} who focused on the relative entropy perturbation. Even if the latter is observationally relevant, a well-defined curvature perturbation requires the vanishing of the total entropy perturbation. \ana{suggest: 
The analysis of \cite{Cicoli:2021yhb} focused on the relative entropy perturbation as the observationally relevant quantity. However a well-defined curvature perturbation requires a well-defined total entropy perturbation, even if it is not directly observable. The total entropy perturbation and its relation to the isocurvature perturbation has also been discussed, in a different context, in \cite{Huston:2011fr,Huston:2013kgl}.}  In~\cite{Gordon:2000hv}, the total entropy perturbation is defined as 
\begin{equation}
\mathcal{S}_{\rm tot} \equiv H \left( {\delta P \over \dot{P}} -  {\delta \rho \over \dot{\rho}} \right) \, ,
\end{equation}
where $\delta\rho$ and $\delta P$ are respectively the perturbations in the energy density and pressure. On super-Hubble scales the curvature perturbation is sourced by the total entropy perturbation as
%which takes the form 
%
\begin{equation}
%\mathcal{R}' \approx 2 \mathcal{S}_{\rm tot} \, ,
\mathcal{R}' = -3 \frac{\dot{P}}{\dot\rho} \mathcal{S}_{\rm tot} 
\, , 
\label{eq:Rprime}
\end{equation}
%
%\oi{clarify $c_s^2$?}
where the prime is differentiation with respect to $e$-folds $N$. %and $c_s^2 \equiv \dot{P}/\dot\rho$.
%Note that in this definition, the total entropy perturbation remains finite as long as $\Omega \, Q_{\rm n}$ does not grow over time.
This implies that any sizable %non-vanishing 
total entropy perturbation will prevent the curvature perturbation from reaching its adiabatic limit and 
$\mathcal R$  will not be conserved after horizon exit \cite{Renaux-Petel:2014htw}. For single-field slow-roll models, $\mathcal{S}_{\rm tot}$ is zero and $\mathcal{R}$ is conserved, consistent with observations on the CMB scales. But this is not the only option; in practice, it is sufficient that $|\mathcal{S}_{\rm tot}| \ll |\mathcal{R}|$ holds a few $e$-folds before the end of inflation {\it and  only for observable scales} \footnote{Transient violations of this condition on smaller scales, even in single-field models, are the basis of various scenarios of primordial black hole formation in the early universe \cite{Byrnes:2021jka,Ozsoy:2023ryl}.} (see e.g.~in shift-symmetric orbital inflation \cite{Achucarro:2019pux}).

Now we turn to the multi-field case. 
The total entropy perturbation is given by 
\begin{equation}
\mathcal{S}_{\rm tot} =  {1 \over 3} \left( {3 H^2 \sigma' \over 2V_{\sigma}} + 1 \right)^{-1} \left[ -{1\over \epsilon} \left( {k \over a H } \right)^2 \Psi  
+ 2 {\Omega \over  \sqrt{2\epsilon}} Q_{\rm n} \right]\, , 
\end{equation}
where we have set $8\pi G=1$, $\sigma$ is the arc length of the trajectory, $\sigma'=\sqrt{2\epsilon}$ with $\epsilon \equiv -\dot{H}/H^2$, and $\Psi$ is the Bardeen potential. Here, $\Omega$ is the turning rate in $e$-folds and $Q_{\rm n}$ is the orthogonal gauge-invariant perturbation, which is the Mukhanov-Sasaki variable projected onto the orthogonal direction to the background field trajectory. Defining 
\begin{equation} 
\label{eq:s_entropy}
s \equiv {\Omega \over  \sqrt{2\epsilon}} Q_{\rm n} \, , 
\end{equation}
on super-Hubble scales we have $\mathcal{S}_{\rm tot} \approx - 2s/3$. From the Einstein equations, we find $\mathcal{R}' \approx 2s$~\cite{Gordon:2000hv}, so, in what follows, we will use $s$ as a proxy for 
the total entropy perturbation on super-horizon scales.
%
%
%
%Then, the total entropy perturbation is written as
%%The total entropy perturbation on superhubble scales obeys the following equation
%%
%\begin{equation} 
%\label{eq:s_entropy}
%s \equiv {\Omega \over  \sqrt{2\epsilon}} Q_{\rm n} \, , 
%\end{equation}
%%
%where $\Omega$ is the turning rate in $e$-folds and $Q_{\rm n}$ is the orthogonal gauge-invariant perturbation, which is the Mukhanov-Sasaki variable projected onto the orthogonal direction to the background field trajectory. 
%
%
%
For definiteness, for the moment we focus on two-field inflation. 
The usual stability analysis focuses on  $Q_{\rm n}$, which obeys the following equation on super-Hubble scales:
\begin{equation}
\label{eq:msuper}
Q_{\rm n}'' + (3 - \epsilon) Q_{\rm n}' + {\mu_{\rm eff}^2 \over H^2} Q_{\rm n} = 0 \, .
\end{equation}
Therefore, the stability of $ Q_{\rm n}$ is determined by the sign of the effective mass $\mu_{\rm eff}^2$. On the other hand, the total entropy perturbation \eqref{eq:s_entropy} obeys
\begin{equation} 
\label{eq:dds}
s'' + (3 - \epsilon + \eta - 2 \eta_{\Omega} ) s' + {M_{\rm s}^2 \over H^2 }s = 0 \, ,
\end{equation}
where $\eta \equiv \epsilon'/\epsilon$ and $\eta_{\Omega}$ is the logarithmic time derivative of the turning rate in $e$-folds: 
\begin{equation}
    \eta_{\Omega} \equiv (\log \Omega)'=\frac{1}{\Omega}\frac{\ud\Omega}{\ud N} \, .
\end{equation}
%
%and  $\epsilon$ and $\eta$ are the slow-roll parameters.
The stability of $s$ is determined by the sign of the \textit{entropy mass} squared $M_{\rm s}^2$ as well as the sign of the ``friction'' term -- the coefficient in front of $s'$.
The entropy mass squared is given by \footnote{Alternatively, defining the radius of curvature of the trajectory $\kappa^{-1} \equiv \Omega/\sqrt{2\epsilon}$ we can express $M_{\rm s}^2$ as
\be
\frac{M_{\rm s}^2}{H^2 }  \equiv {\mu_{\rm eff}^2 \over H^2} + \left[3 - \epsilon + (\log \kappa)' \right] (\log \kappa)' + (\log \kappa)'' \, ,
\ee
where $(\log \kappa)' = \eta/2- \eta_{\Omega}$.}
\begin{equation}
    \frac{M_{\rm s}^2}{H^2 }  \equiv {\mu_{\rm eff}^2 \over H^2}-\eta_{\Omega} \left( 3- \epsilon - \eta_{\Omega} \right) - \eta_{\Omega}' + {1 \over 2}\eta' + {1\over 2} \eta \left({1\over 2}\eta + 3 -\epsilon - 2\eta_{\Omega} \right) \, .
\label{eq:MS}
\end{equation}
Note that, during slow roll inflation, $M_{\rm s}^2\simeq \mu_{\rm eff}^2$ when $\eta_{\Omega} \ll 1$.
In this regime, stability is well captured by the standard analysis based on the sign of $\mu_{\rm eff}^2$. However, 
even if $\mu_{\rm eff}^2$ is negative, $M_{\rm s}^2$ can remain positive if, with $\epsilon$, $\eta\ll1$, the following condition is satisfied:
%
%with $\epsilon$, $\eta\ll1$, a negative effective mass $\mu_{\rm eff}^2$ in $M_{\rm s}^2$ can be compensated by 
%
\begin{equation}
\eta_{\Omega} \lesssim -\mathcal{O}(1) \, .
%\qquad 
%\epsilon,\ \eta \ll 1,
%\qquad
%\text{\textit{(ultra slow-turn)}}
\end{equation}
%
%
%which we dub the regime of \textit{ultra slow-turn} inflation by analogy with the ultra slow-roll regime, where  $\eta \lesssim -\mathcal{O}(1)$. In this regime $M_{\rm s}^2$ becomes positive, while $\mu_{\rm eff}^2$ remains negative. 
This explains the apparent (in)stability paradox seen in the models presented in~\cite{Cicoli:2018ccr,Cicoli:2019ulk,Cicoli:2021yhb,Kallosh:2024kgt,GonzalezQuaglia:2025qem}. We will call this slow-roll regime ``\textit{ultra slow-turn}'' inflation by analogy with the ultra slow-roll regime, where $\eta \lesssim -\mathcal{O}(1)$.

In this paper, we argue that the stability of the total entropy perturbation provides the appropriate criterion for assessing stability and demonstrate its implications. We further present analytical and numerical examples exhibiting ultra slow-turn behavior. Since our analysis focuses on solutions to linear perturbations, it implicitly assumes that the coefficients of the various terms either exhibit weak time dependence, varying slowly in time, or depend strongly on time but decay exponentially fast to zero. Under these assumptions, the eigenvalues of the stability matrix provide a reliable diagnostic for stability.

%This work is organized as follows: \oi{to do}

\section{Stability in the linear theory}
\label{sec:stability}

%\subsection{How to define background stability}
%For a dynamical system $\dot{y}=f(y)$, where $y\equiv \{ \phi^i,\dot{\phi}^i \}$ involves some fields and their derivatives, we are interested in formulating the correct way to study stability. The problem of identifying the correct stability criteria arises because perturbations of the fields $\delta \phi^i$ behave as vectors in the field space. This poses the question of how to quantify the conditions of perturbativity; should we impose $\delta \phi^i \ll \phi_0^i$ for each component or require $||\delta \phi^i||\ll \{\phi_0^i\}$ for all $i$? Irrespective of what is the correct way to define stability, if one can prove that a system is (un)stable in a particular coordinate system, the same should hold in an arbitrary system. It remains challenging to express the conditions for stability in a coordinate-invariant way.

In this section, we address the question of whether the vanishing %effective mass 
of the orthogonal perturbation $Q_{\rm n}$ is the correct way to infer stability unambiguously.  The dynamics of isocurvature perturbations has become the standard method mainly because it correlates well numerically with the existence of attractor and unstable solutions in certain models. Some characteristic examples with strong attractor behaviour include models with non-minimal couplings \cite{Kaiser:2012ak,Kaiser:2013sna} and the rapid-turn attractors \cite{Bjorkmo:2019fls}, while hybrid inflation \cite{Linde:1993cn} or geometrical destabilization \cite{Renaux-Petel:2015mga} belong to the unstable class.\footnote{Note that in both cases -- hybrid and geometrical destabilization/sidetracked -- the {\em specific models} studied have canonical kinetic terms for the field orthogonal to the inflationary trajectory. This evades the question of how to characterize the instability: based on the $L^2$ size of the field perturbation or on the covariant size of the perturbation given by the field space metric. With canonical kinetic terms both are equivalent. 
%This is part of the reason for the apparent paradoxes in the counterexamples and the models considered here.
} 
A common feature of the previous stable examples is a slowly varying turn rate and a very large $\mu_{\rm eff}^2/H^2$ so the difference between $\mu_{\rm eff}$ and $M_{\rm s}$ becomes negligible, as can be read from \eqref{eq:MS}.

Nevertheless, as we already explained in the introduction, counterexamples do exist and, therefore, we can conclude that decaying isocurvature perturbations are not necessary for the existence of a stable solution. 
Defining stability in inflation is a highly nontrivial task; there are at least two issues:

\begin{itemize}

\item  First, as is well-known from single-field models, the inflationary solution is not a fixed point but a {\it trajectory} in phase space, and some standard techniques for dynamical systems are not readily applicable. The issue of time dependence can be alleviated by considering slowly varying quantities such as $\epsilon,\eta,\Omega,$ etc.

\item Second, when formulating stability of multi-field trajectories in terms of distances, it is not clear which choice of measure is the correct one. More specifically, one is forced to choose between an $L^2$-type norm $\sqrt{\sum_i (\delta\phi^i)^2}$ requiring that every orthogonal field perturbation go to zero\footnote{This choice is implicit when we consider eigenvalues of the stability matrix in phase space.} versus the field-space norm of these orthogonal perturbations, $\sqrt{G_{ij}\delta \phi^i\delta \phi^j}$, which implies that the projections along the orthogonal unit vectors go to zero. 
The reason why these norms can give different answers is directly related to the time dependence in $G_{ij}(N) \delta \phi^i\delta \phi^j$ (we will return to this point in Section~\ref{sec:g} for more details).
As we have argued, formulating the criteria in terms of perturbations projected along the kinematic frame does not give the correct result in e.g.~the ultra slow-turn case. 

\end{itemize}

Taking a step back, we observe that our dynamical system, the Klein-Gordon equations for the scalar fields, 
\begin{equation}
\uD_N v^i +(3-\epsilon)\left[v^i + (\log V)^{,i}\right] = 0 \, ,
\end{equation}
involves three vectors as variables: the velocity $v^i \equiv d\phi^i/dN$, the covariant acceleration $a^i \equiv\uD_N v^i$, and the gradient $w^i \equiv  (\log V)^{,i}$, where $D_N = (\phi^j)' \nabla_j = v^j \nabla_j$ is the covariant derivative in e-folds along the trajectory. %Above, $\epsilon$ is the first slow-roll parameter, 
Note that $\epsilon$ can be written in terms of $v^i$ as $\epsilon=G_{ij}v^iv^j/2$.
Instead of formulating stability in terms of fields and their velocities, we can alternatively formulate it in terms of scalar functions constructed from the previous three vectors, and demand that perturbations of these scalars around an attractor solution should decay. Of the six possible scalars that can be constructed from the three vectors, the equations of motion reduce the number of independent scalars to two, which we choose to be $\epsilon$ and $\epsilon_{\rm V}=G^{ij}(\log V)_{,i}(\log V)_{,j}/2$; the rest can be found by taking linear combinations of them and their covariant derivatives.

First, we perturb $\epsilon$:
\begin{align}
\delta \epsilon & = G_{ij}(\phi^i)' \uD_N (\delta \phi^j)'
 \, .
\end{align}
Using $(\phi^i)'=\sqrt{2\epsilon}t^i$ and $\uD_N t^i = \Omega n^i$, where $t^i$ and $n^i$ are tangential and normal vectors to the trajectory, we obtain on super-Hubble scales
\begin{equation}
\label{eq:delta-epsilon}
\delta \epsilon = 2\epsilon \left(  \mathcal{R}'  + {1 \over 2}\eta \mathcal{R} - s \right) = 2\epsilon \left(   {1 \over 2}\eta \mathcal{R} + s \right) 
\, .
\end{equation}
%
%where $\eta=\epsilon'/\epsilon$, and the curvature perturbation is defined in the flat gauge. Above we used $\mathcal{R}' \approx 2s$.\footnote{The total entropy perturbation is given by $\mathcal{S}_{\rm tot} = - {2 \over 3 \epsilon} \left( {3 H^2 \sigma' \over V_{\sigma}} + 2 \right)^{-1} \left( {k \over a H } \right)^2 \Psi  - {2 \over 3 }{\Omega \over  \sqrt{2\epsilon}} Q_{\rm n} \,$, with $8\pi G=1$, $\sigma$ is the arc length, $\sigma'=\sqrt{2\epsilon}$ and $\Psi$ being the Bardeen potential. On super-Hubble scales, it reduces to $\mathcal{S}_{\rm tot}\simeq - {2 \over 3 }s$. Using 
%\eqref{eq:Rprime} we recover $\mathcal{R}' \approx 2s$. }  
%
We observe that the total entropy perturbation $ s  = \Omega Q_{\rm n}/\sqrt{2\epsilon}$ enters \eqref{eq:delta-epsilon}, and not the bare quantity $Q_{\rm n}$. 
If the expression in parenthesis is of order slow-roll, 
perturbations in $\epsilon$ become negligibly small compared to $\epsilon$ and the assumed solution is an attractor.\footnote{Because realistic models of inflation do not have $\eta=0$ exactly on the attractor solution, perturbations in $\epsilon$ can never be exactly zero. For slow-roll models, which is the focus of our work, we measure the smallness of a quantity compared to the characteristic small scale of the problem, i.e.~the slow-roll parameters. } This happens, for instance,
if $\mathcal{R}$ freezes (equivalently if $s \ll \mathcal{R}$) and $\eta \ll 1$.

Next, we study perturbations in the potential slow-roll parameter $\epsilon_{\rm V}$, which reads
\begin{equation}
\delta \epsilon_{\rm V} = (\log V)^{,i} (\log V)_{;ij} \delta \phi^j  \, .
\end{equation}
In the orthonormal frame, the previous expression can be written up to $\mathcal{O}(\epsilon^2,\epsilon \eta, \eta^2,\eta',\eta_{\Omega}')$ corrections (see Appendix~\ref{app:notation})
\begin{align}
\delta \epsilon_{\rm V} 
& = 
\Big[ (\log V)_{\sigma}(\log V)_{\sigma\sigma} + (\log V)_{\rm n}(\log V)_{\rm \sigma n} \Big] Q_{\sigma} 
+ \Big[ (\log V)_{\sigma}(\log V)_{\rm \sigma n} + (\log V)_{\rm n}(\log V)_{\rm n n} \Big] Q_{\rm n} 
%\end{equation}
%%
%which can be written as
%%
%\begin{equation}
\nonumber\\
& =
%\delta \epsilon_{\rm V} = 
\epsilon_V \left\{\left[\eta - 2\left(1  - {\epsilon \over \epsilon_{\rm V}} \right)\eta_{\Omega} \right] \mathcal{R} 
+ {2\epsilon \over \epsilon_V (3-\epsilon)} \left[ 3 -\epsilon - \eta_{\Omega} + {1 \over 2} \eta + (\log V)_{\rm n n} \right] s \right\}
\, .
\end{align}
On a slow-roll attractor solution, the right hand side should be $\mathcal{O}(\text{slow-roll}) \times \mathcal{O}(\epsilon_{\rm V})$. 
%which is true since $\epsilon<\epsilon_{\rm V}$ always holds, and $\epsilon \approx \epsilon_{\rm V}$ for slow-turn models (see equation \eqref{eq:epsvsepsV}), where $\eta_{\Omega}$ can be $\mathcal{O}(1)$, or $\epsilon\ll \epsilon_{\rm V}$ for rapid-turn models with $\eta_{\Omega}=\mathcal{O}(\epsilon,\eta)$.
Once again $s$ and not $Q_{\rm n}$ appears in the perturbations. The previous expressions demonstrate that stability can --and should-- be formulated in terms of $\mathcal{R}$ and $s$. %\oi{discuss with Perseas}

\subsection{Symmetric models} 
\label{sec:g}

To better understand why stability using $Q_{\rm n}$ may fail to give the correct conclusion, we investigate the symmetric models. The common feature of these models is that they have a shift symmetry in the field orthogonal to the inflationary trajectory (the orthogonal field is not canonically normalised, while the inflaton is). As a proxy, let us consider a model where the potential and the metric are given respectively by
\begin{align}
V & = V(\phi) 
\, , 
\\
\ud s^2 & = \ud \phi^2 + g(\phi)^2 \ud \chi^2 \, ,
\end{align}
which is a very good approximation to the modular, SL(2,$\mathbb{Z}$) and the fibre inflation models mentioned in the introduction --where $g$ increases during inflation--, and also to the hyperinflation model \cite{Brown:2017osf} and shift-symmetric $\alpha$-attractors with $\alpha=1/3$ \cite{Kallosh:2015zsa, Achucarro:2017ing} --where $g$ decreases during inflation--. The stability of the slow-roll solution was studied in \cite{Christodoulidis:2019mkj,Christodoulidis:2022vww}, in which it was found that stable inflation along $\phi$ requires 
\begin{equation} 
\label{eq:stab_back}
3-\epsilon + B >0 \, ,
\end{equation}
with
\begin{equation}
B \equiv {\ud \over \ud N}\left( \log  g \right)  =  \left( \log  g \right)_{,\phi} \phi' \, ,
\end{equation}
where they treated $B$ as (approximately) constant on the trajectory $(\log B)' =\mathcal{O}(\epsilon)$. Moreover, the effective mass on super-Hubble scales is given by~\cite{Cicoli:2018ccr,Christodoulidis:2019mkj} 
\begin{equation} 
\label{eq:eff_mass}
{\mu_{\rm eff}^2 \over H^2} = - B (3-\epsilon + B) 
 -  B' \, .
\end{equation}
Note that it is possible to satisfy \eqref{eq:stab_back} with $B>0$ which will make the effective mass negative.
Although the previous expression \eqref{eq:eff_mass} was shown in
\cite{Cicoli:2018ccr,Christodoulidis:2019mkj} to hold approximately, we can use a simple argument to demonstrate full equality based on the shift symmetry of $\chi$. Since the model is symmetric under constant shifts $\tilde{\chi} = \chi + c$ the same should hold for $\delta \chi$. Therefore, the equation for the isocurvature perturbation $Q_{\rm n}$ should also be symmetric under the shift $\tilde{Q}_{\rm n} = Q_{\rm n} + c g$. Substituting this into the equation for $Q_{\rm n}$ and demanding invariance, we arrive at \eqref{eq:eff_mass}.

%If additionally the turn rate becomes vanishingly small, the problem becomes single field with decreasing total entropic perturbation ($s$) and thus no sourcing to the curvature perturbation on superhubble scales. 

But what is $B$ that seems a coordinate-dependent quantity? From the definition of the turning vector $\Omega n^i \equiv \uD_N t^i$ we can derive an exact expression for the turn rate in this class of models:\footnote{This is a special case of the exact relation $3 + \eta_{||} = \Omega \tan\delta$, where $\eta_{||} \equiv t_i \uD_{\rm t} \dot{\phi}^i/(H\dot{\phi})$  and $\delta$ is the angle between $\nabla V$ and the normal vector to the trajectory $n^i$. }
\begin{equation} 
\label{eq:omega_1}
\Omega = \left( 3-\epsilon +{1 \over 2} \eta\right)  \frac{\chi'}{\phi'} g \, .
\end{equation}
As a consistency check, when $\chi'=0$ (radial motion) the turn rate vanishes, as expected. Taking the logarithmic time derivative of eq.~\eqref{eq:omega_1} yields 
\begin{equation} 
\label{eq:log_turn_rate}
\eta_{\Omega}=(\log \Omega)' \approx - \left[3 -\epsilon + (\log g)_{,\phi} \phi' \right] + \mathcal{O}(\epsilon,\eta)\, ,
\end{equation}
which is exactly the quantity appearing in eq.~\eqref{eq:stab_back}. Although the turn rate decreases exponentially, its logarithmic derivative cannot be omitted in the various equations, as it is $\mathcal{O}(1)$. We call this \textit{ultra slow-turn}, akin to the ultra slow-roll scenario where $\epsilon\rightarrow 0$ but $\eta =- \mathcal{O}(1)$ \cite{Tsamis:2003px,Kinney:2005vj}. We illustrate this  behavior with numerical examples in Section \ref{sec:examples}.

Now we examine how the isocurvature perturbation behaves on the previous stable solution where $\chi$ is constant. The equation of motion for $Q_{\rm n}$ \eqref{eq:msuper}
%
%\be \label{eq:msuper}
%Q_{\rm n}'' + (3 - \epsilon)Q_{\rm n}' + \mu_{\rm eff}^2 /H^2Q_{\rm n} = 0 \, , 
%\ee
%
has two solutions $Q_{\rm n} \propto \exp(\lambda_{1,2}N)$ with exponents:
\begin{equation}
\lambda_{1,2} = -{1\over 2} (3-\epsilon) \pm {1\over 2} \sqrt{(3-\epsilon)^2 - 4 \frac{\mu_\text{eff}^2}{H^2}} \, ,
\end{equation}
and the asymptotic behaviour of $Q_{\rm n}$ will be determined from its largest eigenvalue, %which is the one with the plus sign. 
$\lambda_1$. 
Using the expression for the effective mass, we find that the largest eigenvalue is
\begin{equation}
\lambda_{\rm max} = \text{max}\left[ B, -(3-\epsilon+B) \right] \, .
\end{equation}
Depending on the sign of $B$, we consider the following cases:

\begin{itemize}
    \item 
If $B>0$, the maximum eigenvalue is $B$ %, which is consistent with considering $\chi$ as the orthogonal field $Q_{\rm n} \sim \exp(B N) \delta \chi = g\delta \chi$. Therefore, inflation proceeds along $\phi$ (gradient flow) 
and the orthogonal perturbation is exponentially amplified on super-Hubble scales, satisfying $Q_{\rm n} \rightarrow c g$. This is the case for the ultra slow-turn models.

\item If $B<0$, there is a critical value $B_0 \equiv -(3 - \epsilon)$ for which $\lambda_{\max} \leq 0$.  
\begin{itemize}
    \item  For $B_0<B<0$, \eqref{eq:stab_back} is satisfied, and thus the gradient flow solution, $\chi'=0$, is stable. In this case $Q_{\rm n}$ is exponentially suppressed on super-Hubble scales. This happens e.g.~in the `radial' phase of the hyperinflation model or the alpha-attractors with $\alpha=1/3$ for a potential that is not steep enough.

    \item For $B<B_0$, \eqref{eq:stab_back} is not satisfied, and the gradient flow solution becomes unstable. This behaviour is illustrated in the hyperinflation phase (steep potentials).

\end{itemize}
\end{itemize}

A similar qualitative conclusion can be drawn by studying perturbations in Gaussian normal coordinates. Because we have a family of equivalent trajectories, and the geodesic distance between them increases over time, adapting coordinates around a particular one and expanding will result in an apparently unstable solution. More specifically, by applying the transformation
\begin{align}
\phi &= \sigma + {1\over 2} (\log g)_{,\sigma} \chi_{\rm g}^2 \, ,
\\ 
\chi  &= {\chi_{\rm g} \over g} \left[ 1 - {1 \over 3} \chi_{\rm g}^2 (\log g)_{,\sigma} \right] \, ,
\end{align}
the metric and potential around the trajectory admit the form
\begin{align}
\ud s^2 & = \left( 1 + { g_{,\sigma \sigma} \over g} \chi_{\rm g}^2 \right) \ud \sigma^2 + \ud \chi_{\rm g}^2 
+ \mathcal{O}(\chi_{\rm g}^3)  \, , 
\\
V(\sigma,\chi_{\rm g}) & = V(\sigma) + {1\over 2} V_{,\sigma} (\log g)_{, \sigma} \chi_{\rm g}^2 
+ \mathcal{O}(\chi_{\rm g}^3) \, ,
\end{align}
where we can use $\sigma$ and $\phi$ interchangeably for small $\chi_{\rm g}$. The inflationary trajectory is defined on $\chi_{\rm g}=0$ for which the metric becomes canonical. Using the slow-roll approximation $3H \dot{\sigma} \approx -V_{,\sigma}$ the potential can be written as 
\begin{equation} 
\label{eq:pot_fw}
V(\sigma,\chi_{\rm g}) \approx V(\sigma) - {3 H^2\over 2} (\log g)' \chi_{\rm g}^2 + \mathcal{O}(\chi_{\rm g}^3) 
\end{equation}
Recall that the geodesic solution, given by $\chi' = 0$ in the original coordinate system, becomes $\chi_{\rm g} / g$ in Gaussian normal coordinates and approaches a constant for small $\chi_{\rm g}$. Therefore, Gaussian normal coordinates work only if $g$ decreases; if $g$ increases, which is our case of interest, then  
%$x_{\rm g}$ 
$\chi_{\rm g}$ also has to increase to be consistent with the single field solution. The geodesic distance between two adjacent initially parallel trajectories representing close initial conditions, which is roughly equal to $Q_{\rm n}$, will increase over time. This is consistent with the form of the potential shown in eq.~\eqref{eq:pot_fw}. For increasing $g$, the potential features a maximum when evaluated on the $\chi_{\rm g}=0$ solution %and for negative or moderately positive curvature values, 
and the effective mass becomes negative. %Therefore, any Gaussian normal coordinates construction will be unstable. 
Again we stress that, because of the way the Gaussian normal coordinate system is constructed, it can only probe solutions where the geodesic distance between two solutions with close initial conditions decreases.

By contrast, studying the stability using the total entropy perturbation gives the correct answer in all previous cases. For an ultra slow-turn model $\eta_{\Omega} <0$ (equivalently $B>B_0$) the equation for the total entropy perturbation on super-Hubble scales becomes
\begin{equation}
\label{eq:s-ust}
s'' + (3 -\epsilon - 2 \eta_{\Omega})s' - 2 \eta_{\Omega} (3 - \epsilon)s = \mathcal{O}(\eta,\eta_{\Omega}',\eta') \, .
\end{equation}
Since we assumed $\eta_\Omega < 0$ the previous equation has only exponentially decaying solutions with $\mathcal{O}(1)$ exponents. More specifically, ignoring %$\eta',\eta_{\Omega}',\epsilon'$ terms,  
the right-hand side of \eqref{eq:s-ust}
$s$ evolves as 
\begin{equation}
s(N) = c_1 e^{2\eta_{\Omega}N} + c_2 e^{-(3-\epsilon)N} \, ,
\end{equation}
implying a strong suppression of the total entropy perturbation. As a consequence, the curvature perturbation is not sourced on super-Hubble scales and behaves similarly to single-field models. As a final note, the entropy perturbation is related to the $\delta \chi$ field perturbation via
\begin{equation}
s = \Omega/\sqrt{2\epsilon} g \delta \chi \approx e^{-3N} \delta \chi \, ,
\end{equation}
where we used \eqref{eq:log_turn_rate} to relate $\Omega$ to $g$.

\subsection{When is $\mu_{\rm eff}^2<0$ an instability?}

To study more explicitly the stability of an inflationary solution, we rewrite the equations for the two perturbations (see e.g.~\cite{Christodoulidis:2023eiw}) as
\begin{align}
&\mathcal{R}'  = p_\mathcal{R} + 2 s \, , 
\\ 
& p_\mathcal{R}' + (3 - \epsilon +\eta)p_\mathcal{R} - {\nabla^2 \over a^2} \mathcal{R} = 0 \, , 
\\
&s'' + \left(3 - \epsilon + \eta  -2 \eta_{\Omega} \right) s ' + \left( {M_{\rm s}^2 \over H^2} - {\nabla^2 \over a^2} \right) s  =   - 2 p_\mathcal{R}  \, , 
\end{align}
where we used the ``canonical momentum'' $p_{\mathcal{R}}$, and $M_{\rm s}^2$ is as defined \eqref{eq:MS}. 
%
%and we reproduce it here for convenience
%\begin{equation}
%    \frac{M_{\rm s}^2}{H^2 }  \equiv {\mu_{\rm eff}^2 \over H^2}-\eta_{\Omega} \left( 3- \epsilon - \eta_{\Omega} \right) - \eta_{\Omega}' + {1 \over 2}\eta' + {1\over 2} \eta \left({1\over 2}\eta + 3 -\epsilon - 2\eta_{\Omega} \right) \, .
%\end{equation}
%
 To study the eigenvalues of this system we assume the masses to be adiabatically evolving and hence prime derivatives of various quantities should be suppressed; up to corrections in $\eta' -2\eta_{\Omega}'\ll1$ %and $|\eta|\ll 1$ 
 we find the four (approximate) eigenvalues, $\lambda_1=0$, $\lambda_2 =-(3-\epsilon + \eta )$ and the pair 
\begin{align} 
\label{eq:eigenvalues}
\lambda_{3,4} = -{1\over 2} (3-\epsilon +\eta -2\eta_{\Omega}) \pm  {1\over 2} \sqrt{(3-\epsilon + \eta -2\eta_{\Omega})^2 - 4 \frac{M_{\rm s}^2}{H^2}} \, . 
\end{align}
Demanding a decaying entropy perturbation $s$ and a constant $\mathcal{R}$ we find the following conditions for stability:
%
%\begin{mdframed}
\begin{equation} 
\label{eq:stability_conditions}
M^2_{\rm s} >0 \qquad \text{and} \qquad 3 - \epsilon + \eta -2 \eta_\Omega > 0 \, .
\end{equation}
%\end{mdframed}
%
Note that conditions \eqref{eq:stability_conditions} are manifestly coordinate independent and reduce to the conditions found in \cite{Christodoulidis:2019mkj,Christodoulidis:2022vww} for the assumed geometries. They are the most general conditions that guarantee stability, including all typically encountered cases in the literature, such as $\mu_{\rm eff}^2>0$ with slow-roll conditions $\epsilon,|\eta|,|\eta_{\Omega}| \ll 1$ which cover both the typical slow-roll-slow-turn case but also the slow-roll-rapid-turn attractors.
For the ultra slow-turn models, $\mu_{\rm eff}^2<0$ is compensated by the negative $\eta_{\Omega}\simeq -\mathcal{O}(1) $, while $M^2_{\rm s} >0$ remains positive. The case $M_{\rm s}=0$ is interesting, we defer its discussion to  Section~\ref{sec:massless}.

%There is a fundamental difference between ultra-slow-turn models and other models in works that discuss the effect of tachyonic isocurvature perturbations, for instance, \cite{Linde:1993cn,Gong:2011uw,Renaux-Petel:2015mga}. There, the moment the effective mass becomes negative, the background also gets destabilized, indicating a true instability. In these models, the turn rate starts to increase, leading to $\eta_{\Omega}>0$ violating the second condition in \eqref{eq:stability_conditions}. The total entropy perturbation starts to increase, feeding $\mathcal{R}$ and thus destabilizing the background trajectory until the system falls into the true attractor.

%\textbf{Discuss timescale of convergence/instability}

Finally, we discuss the case where $M_{\rm s}^2<0 $ or $3-\epsilon + \eta - 2\eta_\Omega < 0$. In this case, $s$ will increase exponentially over time, leading to a destabilization of $\mathcal{R}$. However, the timescale for this instability depends on the initial value of $s$; unlike $\mathcal{R}$, whose amplitude is set by the amplitude of quantum fluctuations, $s$ crucially depends on the initial value of the turn rate. We can obtain a rough estimate of the timescale of the instability by assuming that $s$ will significantly source $\mathcal{R}$ when %it becomes of the same order. 
$s \sim \mathcal{R}$. 
Therefore, given an initial value $s_0$, the destabilization of the background will occur after a given number of $e$-folds given by
\begin{equation}
\Delta N_{\rm inst} \sim \lambda^{-1} \left(\log \mathcal{R} - \log s_0 \right) \sim - \lambda^{-1} \log \Omega_0 \, .
\end{equation}
Note that sub-leading contributions were neglected in the eigenvalues in \eqref{eq:eigenvalues}, which can become important especially when the instability is mild. 
A more precise computation can be done via the WKB approximation.

%\textbf{relocate it!}

There is the possibility that  $\mu_{\rm eff}^2 > 0 $ but the system is unstable  because either of the conditions in \eqref{eq:stability_conditions} is violated.  This can happen if $\Omega/\sqrt{2\epsilon}$ increases faster than $Q_{\rm n}$ reaches zero and comparing the two criteria in \eqref{eq:stability_conditions} we can conclude that this necessarily happens when $2\eta_{\Omega} - \eta >3$. In this case, both $\mathcal{R}$ and $s$ increase over time, so we conclude the system is unstable. Looking at only the sign of $\mu_{\rm eff}^2$ would lead to a wrong conclusion about stability. We are not aware of any model in the literature that follows this behaviour.

%We can illustrate this with a recently presented model of modular inflation \cite{GonzalezQuaglia:2025qem}. \textbf{Rephrase}
%\section{Connection to the literature}
%\section{Characteristic examples}
%\subsection{Sidetracked and other unstable models}
%\pc{Oksana do a simulation for sidetracked inflation and show that either or both of the previous two conditions are violated when departing from the slow-roll solution to the rapid-turn attractor.}
%\subsection{$SL(2,{\mathbb{Z}})$ cosmological Attractors and Fibre inflation}
%These have an approximate shift symmetry on the orthogonal field.
%\ana{@Oksana: do you have the simulations showing that sidetracked gets destabilized when $M_s^2$ (not $\mu^2$) becomes negative?}
%\subsection{Modular inflation} \label{sec:modular}
%For illustrative purposes we use the modular inflation model recently introduced in \cite{GonzalezQuaglia:2025qem}. 

\subsection{Examples}\label{sec:examples}

\begin{figure*}
\begin{center}
\includegraphics[scale=0.54]{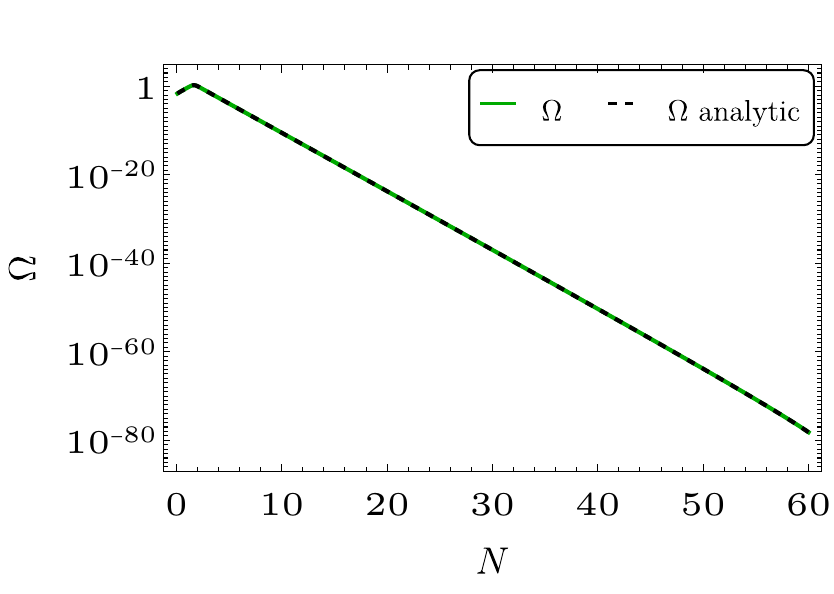}
\includegraphics[scale=0.52]{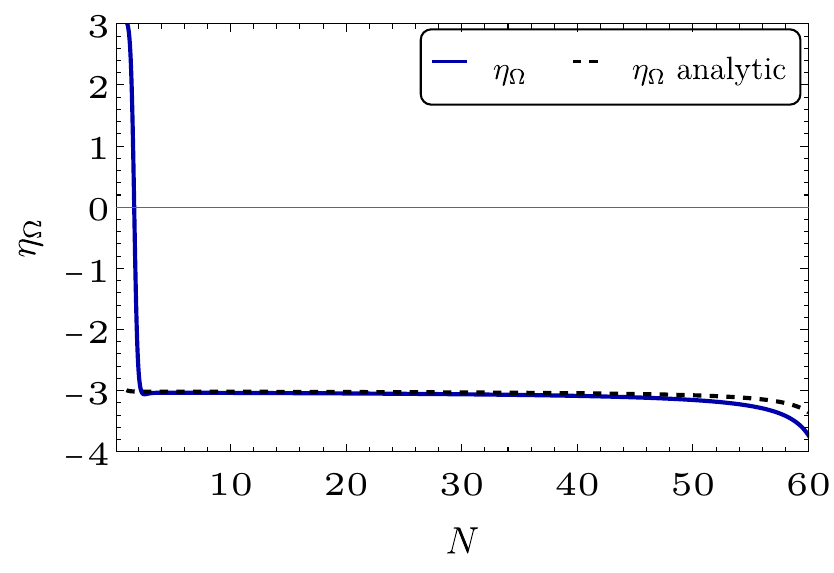} 
\end{center}
\caption[]{
Demonstration of the ultra slow-turn regime: evolution of the turn-rate (left) and its logarithmic time derivative (right) as functions of the number of e-folds $N$ for SL(2,$\mathbb{Z}$) model \eqref{eq:VSL2Z} and \eqref{eq:GSL2Z}. Solid curves correspond to the full numerical solution, while dashed curves represent the analytical expressions given in \eqref{eq:omega_1} and \eqref{eq:log_turn_rate}, with excellent agreement between the two. 
The parameter values we use are $\alpha=1/3$, $\beta=12^3$, $\phi_0=3.3$, $\chi_0=0.3$, $\phi'_0=1$, $\chi'_0=1/g(\phi_0)$.
}
\label{Fig:OmegaEtaSL2Z}
\end{figure*}

To illustrate the ultra slow-turn behavior, we use the SL(2,$\mathbb{Z}$) model, introduced in \cite{Kallosh:2024ymt}, that falls into the class of symmetric models. We consider the following form of the potential and the field-space metric, explored in \cite{Carrasco:2025rud}:
\begin{align}
    V&=V_0\left(1-\frac{\log \beta}{2\pi}\,e^{-\sqrt{\frac{2}{3\alpha}}\phi} \right) \, , 
    \label{eq:VSL2Z}
%\end{equation}
%%
%and the field-space metric
%%
%\begin{equation} 
    \\
    \ud  s^2 & =\ud \phi^2
+\frac{3\alpha}{2}e^{-2\sqrt{\frac{2}{3\alpha}}\phi} \ud\chi^2 \, . 
   \label{eq:GSL2Z}
\end{align}
Figure~\ref{Fig:OmegaEtaSL2Z} shows the evolution of the turn-rate $\Omega$ and its logarithmic time derivative $\eta_{\Omega}$. The turn rate exhibits an exponential decay, while $\eta_{\Omega}\simeq -3$ for this class of models,\footnote{We have also checked that $\Omega$ and $\eta_\Omega$ for the Starobinsky potential $V\propto \left(1-e^{-\sqrt{\frac{2}{3\alpha}}\phi} \right)^2 $ with the same field-space metric, leads to the very similar behavior.} and therefore cannot be neglected. The numerical results show very good agreement with the analytical solutions given in \eqref{eq:omega_1} and \eqref{eq:log_turn_rate}. %The parameter values used for Figure \ref{Fig:OmegaEtaSL2Z} are $\alpha=1/3$, $\beta=12^3$, $\phi_0=3.3$, $\chi_0=0.3$, $\phi'_0=1$, $\chi'_0=1/g(\phi_0)$.

\begin{figure*}
\begin{center}
\includegraphics[scale=0.53]{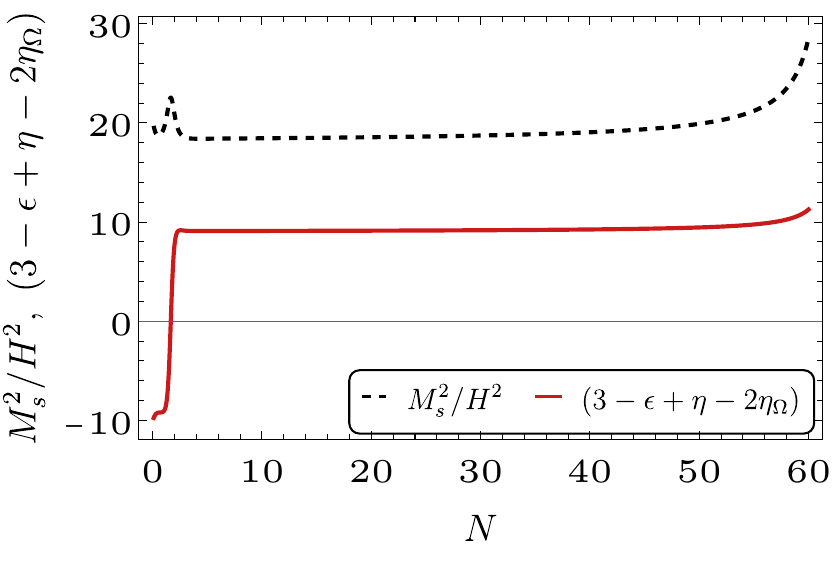} 
\includegraphics[scale=0.52]{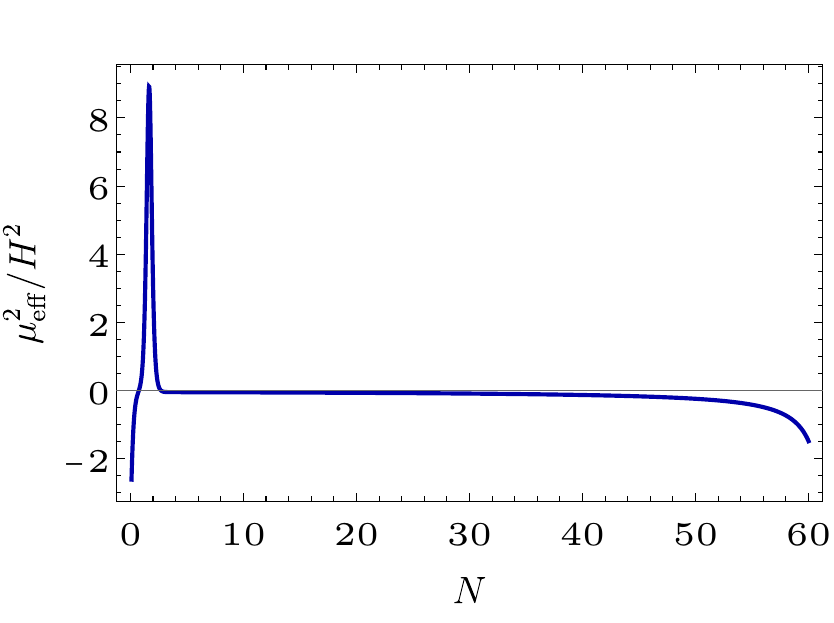}
\end{center}
\caption[]{The two stability conditions (left) from \eqref{eq:stability_conditions} and the effective isocurvature mass on superhorizon scales $\mu^2_{\rm eff}/H^2$ (right) for the SL(2,$\mathbb{Z}$) model for the same parameters as in Figure \ref{Fig:OmegaEtaSL2Z}. Even though $\mu^2_{\rm eff}/H^2<0$, both stability conditions remain positive, implying the stability of the perturbations.}
\label{Fig:MSMuSL2Z}
\end{figure*}

To demonstrate the stability conditions, we again consider the SL(2,$\mathbb{Z}$) example. In Figure \ref{Fig:MSMuSL2Z} we show the evolution of $M_s^2/H^2$ and $3-\epsilon+\eta-2\eta_{\Omega}$, as defined in \eqref{eq:stability_conditions}, together with the effective mass $\mu^2_{\rm eff}/H^2$. Although the effective mass $\mu^2_{\rm eff}/H^2<0$ is negative, the two stability conditions remain positive, indicating stable perturbations.

\subsubsection{Long-lived instability}

\begin{figure*}
\begin{center}
\includegraphics[scale=0.53]{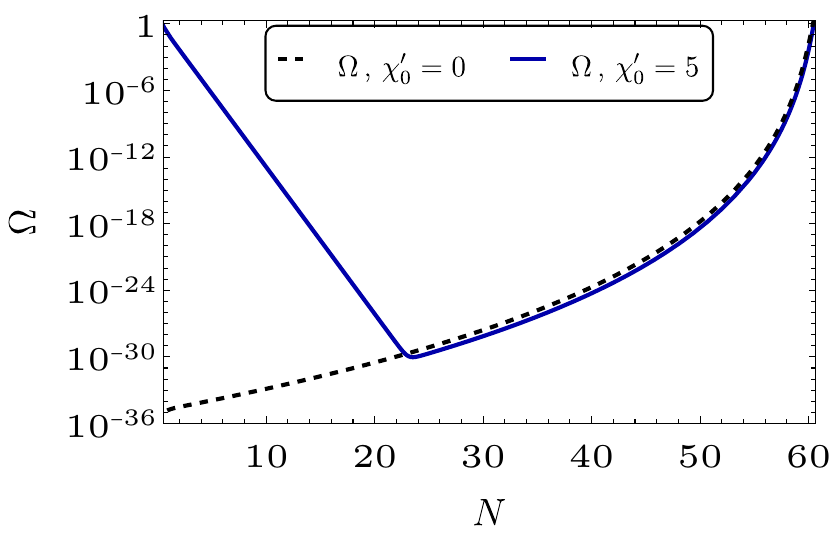} 
\includegraphics[scale=0.53]{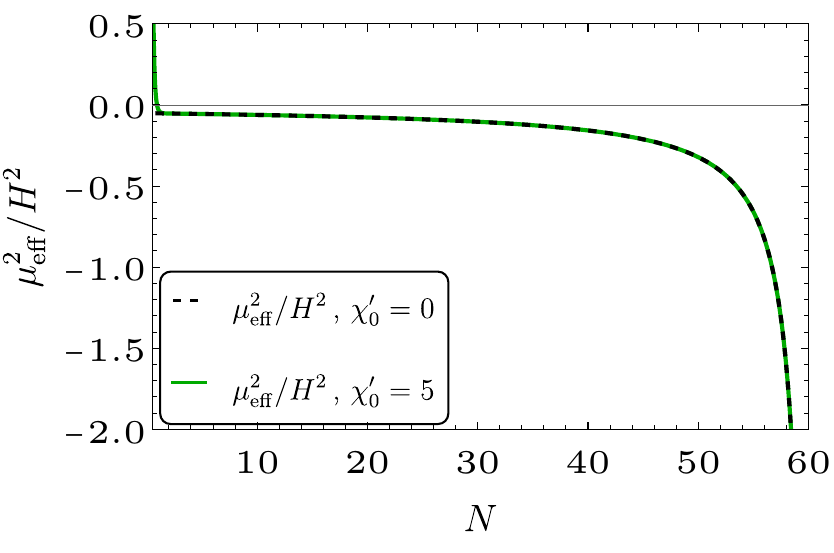} 
\end{center}
\caption[]{The turn-rate $\Omega$ (left) and  $\mu^2_{\rm eff}/H^2$ (right) in the modular inflation model as functions of the number of e-folds $N$, shown for zero initial angular velocity $\chi'_0 = 0$ (black dashed) and for $\chi'_0 = 5$ (solid blue and solid green). A nonzero initial angular velocity qualitatively alters the evolution of the turn rate. The effective mass $\mu^2_{\rm eff}/H^2$ shown in the right panel remains largely insensitive to this change due to the extremely small magnitude of the turn rate, except at early times. Consequently, the curves are nearly indistinguishable. We have set $\alpha=1/3$ and $\beta=1/5$, $\phi_0=4.35$, $\chi_0=0.3$ and $\phi'_0=0$.}
\label{Fig:OmegaModular}
\end{figure*}

\begin{figure*}
\begin{center}
\includegraphics[scale=0.53]{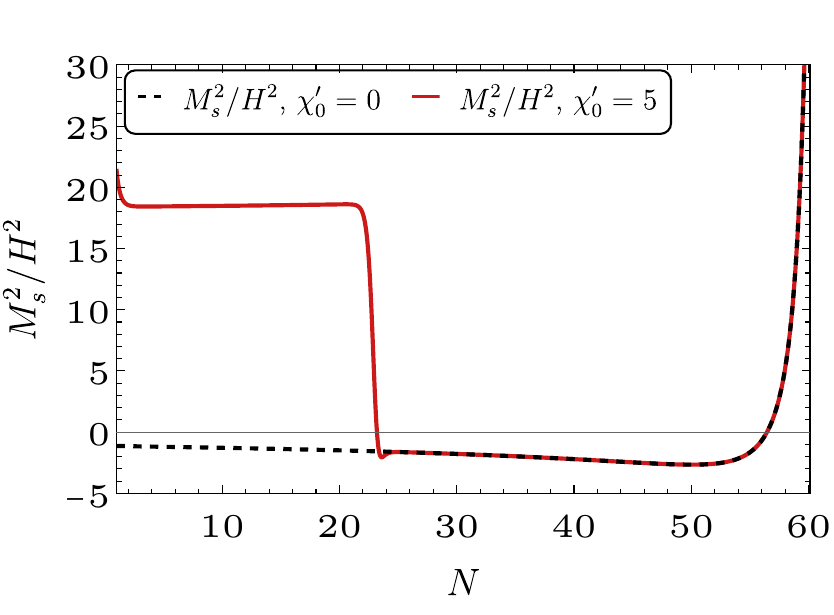}
\includegraphics[scale=0.53]{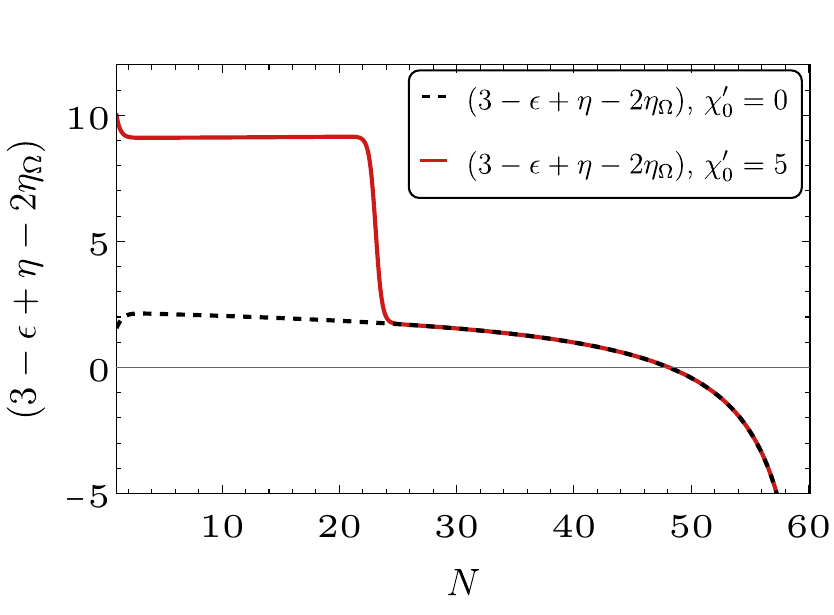} 
\end{center}\caption[]{The stability conditions \eqref{eq:stability_conditions} for the modular inflation model, shown for zero initial angular velocity $\chi'_0 = 0$ (dashed black curves) and $\chi'_0 = 5$ (solid red). Both conditions fail at the end of inflation, $M_s^2/H^2<0$ and $(3-\epsilon+\eta-2\eta_{\Omega})<0$, indicating that the system is unstable. }\label{Fig:MScondMod}\end{figure*}

An interesting example of an unstable situation (that is, however, sufficiently long-lived to support inflation) is a toy model introduced in \cite{GonzalezQuaglia:2025qem} as an approximation to the modular inflation model, mimicking its most relevant features. It has the field-space metric \eqref{eq:GSL2Z} and the hyperbolic potential: 
\begin{equation}
    V(\phi, \chi)=V_0\left[\tanh^2\left( \frac{\phi}{\sqrt{6\alpha}}\right) +\beta^2 e^{-2\sqrt{\frac{2}{3\alpha}\phi}}e^{-\frac{4}{3\alpha}\phi^2}\cos^2(\pi \chi)\right] \, .
\end{equation}
%
%with $\alpha=1/3$ and $\beta=1/5$, $\phi_0=4.35$, $\chi_0=0.3$, $\phi'_0=0$. 
In Figure \ref{Fig:OmegaModular} we demonstrate the evolution of the turn-rate and the effective mass $\mu^2_{\rm eff}/H^2$. The case with zero initial velocity $\chi'_0=0$ agrees with the results presented in \cite{GonzalezQuaglia:2025qem}.
We find, however, that introducing a nonzero initial velocity $\chi'_0\neq0$ modifies the background evolution, causing an early decrease in the turn-rate. It is interesting to note that, in this case, the evolution of the system begins in the rapid-turn regime, evolves toward the ultra slow-turn attractor, and eventually returns to the rapid-turn evolution (non-slow roll). This could have intriguing consequences for non-Gaussianity in this class of models (see for instance \cite{Iarygina:2023msy}), leaving the CMB predictions intact. The effective mass is largely insensitive to this change throughout the evolution except the very initial moments, since the turn rate remains very small. 
Figure \ref{Fig:MScondMod} shows the two stability conditions for cases with zero and non-zero initial velocities $\chi'_0$. For the case with zero initial velocity, $M_s^2/H^2<0$ is negative, indicating the instability in the system. The second condition $(3-\epsilon+\eta-2\eta_{\Omega})>0$ remains positive closely until the end of inflation. Although the modular inflation model appears formally unstable, the turn rate is extremely small ($\Omega\sim 10^{-25}$) throughout the evolution, effectively preventing any destabilization during inflation.

%For this model, the field-space metric has the form
%\begin{equation}
%    ds^2=(\partial\varphi)^2
%+\frac{3\alpha}{2}e^{-2\sqrt{\frac{2}{3\alpha}\varphi}}(\partial\theta)^2, \end{equation}
%with the hyperbolic potential
%\begin{equation}
%    V(\varphi, \theta)=V_0\left(\tanh^2\left( \frac{\varphi}{\sqrt{6\alpha}}\right) +\beta^2 e^{-2\sqrt{\frac{2}{3\alpha}\varphi}}e^{-\frac{4}{3\alpha}\varphi^2}\cos^2(\pi \theta)\right).
%\end{equation}
%In our simulations, we use $\alpha=1/3$ and $\beta=1/5$. We show the evolution in Figures~\ref{Fig:Omega} and \ref{Fig:MScond}.

%\begin{figure*}
%\begin{center}
%\includegraphics[scale=0.55]{figures/phitheta.pdf} 
%\hspace{1.2cm} \includegraphics[scale=0.55]{figures/3D.pdf}
%\includegraphics[scale=0.55]{figures/Omega.pdf}
%\includegraphics[scale=0.545]{figures/mu.pdf} 
%\includegraphics[scale=0.54]{figures/Ms.pdf}
%\includegraphics[scale=0.54]{figures/condition2.pdf} 
%\includegraphics[scale=0.535]{figures/EtaOmega.pdf}
%\end{center}\caption[]{Evolution of the background quantities $\varphi$, $\theta$, turn-rate $\Omega$, effective masses $\mu^2/H^2$ and $M_S^2/H^2$ for modular inflation model with different initial angular velocities. In both cases, we set $\varphi_0 = 3.5$, $\theta_0 = 0.01$, and $\varphi'_0 = 0$. The angular velocities are $\theta'_0 = 0$ (blue dashed curves) and $\theta'_0 = 5$ (green solid curves), respectively. The top left panel shows the background evolution, with the brown curves representing the evolution of $\theta$, while the $\varphi$ curves (blue dashed and solid green) closely overlap.}\label{Fig:modular}\end{figure*}

%\begin{figure*}
%\begin{center}
%\includegraphics[scale=0.535]{figures/MsCondtheta0v2.pdf}
%\includegraphics[scale=0.55]{figures/MsCondtheta5v2.pdf} 
%\end{center}\caption[]{Evolution of the effective mass $M^2_{\rm s}$ (dashed black curves) and the stability condition defined in %\eqref{eq:stability_conditions} (solid red) for zero initial angular velocity $\theta'_0 = 0$ (left panel) and for $\theta'_0 = 5$ 
%(right panel).}\label{Fig:MScond}\end{figure*}

\section{Massless total entropy perturbation} 
\label{sec:massless}

Interestingly, the limit where the entropy mass $M_{\rm s}$ becomes zero is the ultra-light scenario in which the curvature perturbation increases linearly with $N$ \cite{Achucarro:2016fby}.\footnote{In shift-symmetric orbital inflation the isocurvature perturbation is exactly massless due to a shift symmetry in the action for perturbations. In some cases this is related to a symmetry of the equations of motion, or to a scaling symmetry of the action \cite{Achucarro:2019lgo}.} 
This exemplifies the case where $|s|\ll |\mathcal{R}|$ which is definitely the case after many $e$-folds, i.e. on the largest scales just before the end of inflation. In this case, the linear stability analysis is inconclusive due to the existence of a second zero eigenvalue and a Jordan block with an off-diagonal element. %However, the background could still be stable depending on the higher-order terms, and conclusive argument requires center manifold techniques. If this happens, the system is non-perturbative, and a reliable calculation of the curvature perturbation requires lattice simulations.  

%\textbf{Edit-elaborate on ultra-light} an interesting example where the total entropy perturbation freezes on superhorizon scales  but curvature perturbations are approximately adiabatic on CMB scales is

%We use the superpotential method to generalize the previous solutions. 
Another example of massless total entropy perturbation with $\eta_{\Omega} = \mathcal{O}(1)$ can be found as follows. For simplicity we consider a diagonal metric
\begin{equation}
\ud s^2 = f(\chi)^2 \ud \phi^2 + g(\phi)^2 \ud \chi^2 \, ,
\end{equation}
and analyze the solution where $\phi$ is evolving and $\chi' = 0$. Imposing the latter requires the vanishing of the effective gradient
\begin{equation} 
\label{eq:eff_gradient}
{1 \over g^2}\Big[ (3 -\epsilon)(\log V)_{,\chi} - (\log f)_{,\chi} 2\epsilon \Big] = 0 \, ,
\end{equation}
which can be checked using the equation of motion for $\chi$. In the ultra slow-turn models of Section~\ref{sec:g} the previous is identically satisfied because $f_{,\chi} = V_{,\chi} = 0$ for all $\chi$, % leading to a family of such solutions, 
and the entropy perturbation is massive. To obtain a range of solutions, one option for the effective gradient is to vanish identically in a $\chi$-interval, which, however, leads us back to the shift-symmetric orbital models.

A rapidly increasing metric function $g$ will dynamically suppress the effective gradient \eqref{eq:eff_gradient} 
%(assuming that the terms in parentheses do not increase as fast as $g$) 
without requiring $\chi$ to solve the previous equation. This would allow $\chi$ to effectively freeze at an arbitrary value that mimics shift-symmetric orbital models. Since this value is not a critical point of the potential, the turn rate is found as
\begin{equation}
\Omega = -{{3 - \epsilon} \over g\sqrt{2\epsilon}} (\log V)_{,\chi} \, ,
\end{equation}
and if every quantity except $g$ has a weak dependence on time (as in most conventional slow-roll models) we find
\begin{equation} 
\label{eq:eta_omega_g}
\eta_{\Omega} \approx - (\log g)' \, .
\end{equation}
Because we have assumed  $(\log g)'>0$ these models necessarily have $\mu_{\rm eff}^2<0$, as we have explained in Section~\ref{sec:g}. Moreover, since we study the solution where the normal vector coincides with the $\chi$-direction, the isocurvature perturbation is $Q_{\rm n} = g \delta \chi$. Using~\eqref{eq:eta_omega_g} we find that $s \approx \delta \chi$ and so from~\eqref{eq:eff_mass} the mass term of the $\delta \chi$ equation would coincide with $M_{\rm s}$ and we find
\begin{equation}
{\mu^2_{\rm eff} \over H^2} - \eta_{\Omega}(3-\epsilon - \eta_{\Omega}) + \eta_\Omega' \approx 0 \, , %\Rightarrow M^2_{\rm s} = \mathcal{O}(\epsilon,\eta,\eta') \, ,
\end{equation}
leading to $M^2_{\rm s} = \mathcal{O}(\epsilon,\eta,\eta')$,
which demonstrates the cancellation between the effective mass and $\eta_{\Omega}$ leading to massless total entropy perturbation.

\section{Summary and discussion}
\label{sec:summary}

Motivated by recent examples of viable inflationary models with tachyonic isocurvature perturbations, we have revisited the stability of such systems. We have advocated for the use of the total entropy perturbation as a diagnostic tool for stability in slow-roll multi-field models and the stability criteria are summarized in eq.~\eqref{eq:stability_conditions}. We have shown that all known stable examples with tachyonic isocurvature perturbations fall in a class of models that we called ultra slow-turn where an exponentially decreasing turn rate generally causes the total entropy perturbation to decay and one can still recover single-field behaviour on the largest scales. In the shift-symmetric limit, the apparent instability of the isocurvature perturbation is related to the existence of a family of equivalent trajectories, whose distance increases during inflation. We also note that, for this class of models, the ultra slow-turn phase is an attractor of the background dynamics and does not require fine-tuned initial conditions.

%{\color{red}{We leave non-gaussianities for future work.}}

While the dynamics is recovered to be single-field, still non-linear interactions are interesting as genuine multi-field effects due to the entropy perturbation become manifest. Especially, the three-point interaction between two adiabatic and one entropy perturbations is solely boosted by $\eta_\Omega$. This possibly gives rise to large cross-bispectrum with distinguishable shapes. We leave a detailed study on non-Gaussianity for future work.

\section*{Acknowledgements}
We thank Sebastian C\'espedes, Diederik Roest and Dong-Gang Wang for discussions.
AA's work is partially supperted by the Netherlands Organization for Scientific Research (N.W.O.). PC and JG are supported in
part by the Basic Science Research Program through the National Research Foundation of Korea (RS-2024-00336507). The work of OI is supported by the European Union's Horizon 2020
research and innovation program under the Marie Skłodowska-Curie grant
agreement No.~101106874, and by VR Starting Grant 2025-04140 of the Swedish Research Council. 
AA, PC and JG are grateful to the Institut Pascal for hospitality during the workshop ``Cosmology Beyond the Analytic Lamppost'' where parts of this work have been discussed and presented.
JG thanks the Asia Pacific Center for Theoretical Physics for hospitality while this work was under progress.
%
%OI is grateful to Universiteit Leiden for hospitality, where parts of this work have been completed.

\appendix

\section{Notation and useful expressions} 
\label{app:notation}

In the following, we summarize exact expressions valid for any number of fields (see e.g.~\cite{Christodoulidis:2022vww}) that we use in the main text:
\begin{align}
\epsilon_V &= \epsilon \left[  \left( 1 + {\eta \over 2(3-\epsilon)} \right)^2  + \left({\Omega \over 3-\epsilon} \right)^2 \right] \label{eq:epsvsepsV} \\
(\log V)_{\sigma} &\equiv t^i\nabla_i(\log V) =  -\sqrt{2\epsilon} \frac{3-\epsilon + \eta/2}{3-\epsilon} \, , \\
(\log V)_{\rm n} &\equiv n^i\nabla_i(\log V)  =  -\sqrt{2\epsilon}{\Omega \over 3-\epsilon} \, ,\\
(\log V)_{\rm n \sigma} &\equiv t^in^j\nabla_i\nabla_j(\log V) = - \Omega  \left( 1 + {\eta \over 3- \epsilon} - {\eta_{\Omega} \over 3- \epsilon} - {\epsilon \eta \over (3- \epsilon)^2 } \right)  \, , \\
(\log V)_{\sigma \sigma} &\equiv t^it^j\nabla_i\nabla_j(\log V) = { \Omega^2 \over 3- \epsilon}  - {1 \over 2} \eta - {\eta' \over 2(3 - \epsilon)} -  {\eta^2 \over 4(3 - \epsilon)} -  {\epsilon \eta^2 \over 2(3 - \epsilon)^2}  \, .
\end{align}

\bibliographystyle{JHEP}
\bibliography{refs.bib}

\end{document}